\documentclass[twocolumn,aps,pra,amsmath,amssymb,notitlepage]{revtex4-1}
\usepackage[latin9]{inputenc}
\usepackage{mathrsfs}
\usepackage{amsmath}
\usepackage{amssymb}
\usepackage{graphicx}
\usepackage{esint}

\makeatletter

\newcommand*\LyXThinSpace{\,\hspace{0pt}}

\usepackage{epsfig}
\usepackage{bm}
\usepackage{dcolumn}
\usepackage{color}

\makeatother

\begin{document}
\title{Photo-excitation measurement of Tan's contact for a strongly interacting
Fermi gas}
\author{Jia Wang$^{1}$, Xia-Ji Liu$^{1}$, and Hui Hu$^{1}$}
\affiliation{$^{1}$Centre for Quantum Technology Theory, Swinburne University
of Technology, Melbourne, Victoria 3122, Australia}
\date{\today}
\begin{abstract}
We derive theoretically an exact relation between Tan's universal
contact and the photo-excitation rate of a strongly interacting Fermi
gas, in the case of optically transferring fermionic pairs to a more
tightly bound molecular state. Our deviation generalizes the relation
between Tan's contact and the closed-channel molecular fraction found
earlier by Werner, Tarruell and Castin in Eur. Phys. J. B \textbf{68},
401 (2009). We use the relation to understand the recent low-temperature
photo-excitation measurement in a strongly interacting $^{6}$Li Fermi
gas {[}Liu \textit{et al.}, arXiv:1903.12321{]} and show that there
is a reasonable agreement between theory and experiment close to the
unitary limit. We propose that our relation can be applied to accurately
measure Tan's contact coefficient at finite temperature in future
experiments.
\end{abstract}
\maketitle

\section{Introduction}

In 2008, in a series of seminal works \citep{Tan2008a,Tan2008b,Tan2008c},
Shina Tan presented a set of elegant exact relations, showing that
the short-range, large-momentum and high-energy behaviors of a strongly
correlated atomic gas, both statically and dynamically, can be universally
governed by a coefficient. These relations together with the coefficient,
namely Tan's relations and Tan's contact coefficient, pave an entirely
new direction to understand complicated quantum many-body systems
\citep{Braaten2008,Zhang2009,Combescot2009}.

To date, Tan relations have been experimentally verified in both Fermi
gases \citep{Stewart2010,Kuhnle2010} and Bose gases \citep{Wild2012}.
The key Tan's contact coefficient has also been measured in a number
of ways \citep{Stewart2010,Kuhnle2010,Sagi2012,Hoinka2013,Carcy2019,Mukherjee2019},
following different Tan relations \citep{Punk2007,Werner2009,Braaten2010,Taylor2010,Son2010,Hu2010,Hu2011,Hu2012}.
In particular, great efforts have been taken to \emph{accurately}
determine the contact coefficient of a unitary Fermi gas with an infinitely
large $s$-wave scattering length, by using Bragg spectroscopy \citep{Hoinka2013,Carcy2019}
and radio-frequency (rf) spectroscopy \citep{Mukherjee2019}. At finite
temperature, the measured contact of a homogeneous unitary Fermi gas
clearly shows a dramatic change at the superfluid transition temperature,
providing an unambiguous signature for the onset of the superfluid
transition \citep{Carcy2019,Mukherjee2019}.

In this work, we would like to propose that photo-excitation, in which
a fermionic pair absorb a photon to form an excited molecule, may
give an alternative, potentially more accurate method to measure finite-temperature
Tan's contact for a strongly interacting Fermi gas at the crossover
from a Bose-Einstein condensate (BEC) of tightly bound dimers to a
Bardeen-Cooper-Schrieffer (BCS) superfluid of loosely bound Cooper
pairs \citep{Giorgini2008,Randeria2014}. This opens the way to experimentally
determine the superfluid transition temperature at the entire BEC-BCS
crossover, which remains elusive so far.

Our proposal is based on the derivation of a new exact Tan relation
for the photo-excitation rate, if we optically transfer fermionic
pairs to a more tightly bound molecular state \citep{Partridge2005,Paintner2019,Liu2019arXiv}.
In the limit of a weak laser intensity or a small Rabi coupling, our
relation recovers the known relation between Tan's contact and the
closed-channel molecular fraction, predicted earlier by Werner, Tarruell
and Castin \citep{Werner2009}. The advantage of our new relation
is that it also works at a moderately large laser intensity, where
the accuracy of the photo-excitation measurement could be greatly
enhanced. This leads to a more accurate way to measure Tan's contact
at finite temperature. As an application, we use our relation to better
understand a recent photo-excitation measurement in a strongly interacting
$^{6}$Li Fermi gas at the University of Science and Technology of
China (USTC), Shanghai \citep{Liu2019arXiv}. We show that the puzzling
huge discrepancy between theory and experiment, at about two orders
of magnitude, can be reasonably resolved, although there is a still
a factor of three difference found on the BCS side at the largest
magnetic field considered in the experiment.

The rest of the paper is organized as follows. In the next chapter
(Sec. II), we sketch the photo-excitation scenario and present the
model Hamiltonian. In Sec. III, we derive the Tan relation for the
photo-excitation rate. In Sec. IV, we discuss the experimental relevance,
first on analyzing the recent photo-excitation measurement and then
on the proposal for future finite-temperature contact measurements.
Finally, we conclude in Sec. V. 

\section{Model Hamiltonian}

We consider the photo-excitation experiments at Rice \citep{Partridge2005},
Ulm \citep{Paintner2019}, and USTC Shanghai \citep{Liu2019arXiv},
all of which use a strongly interacting $^{6}$Li Fermi gas near the
Feshbach resonance at $B_{0}\simeq832.18$ G, where fermionic pairs
are in the superposition with a stable, ground molecular bound state
$X^{1}\Sigma_{g}^{+}(v=38)$. A resonant laser transition is used
to photo-excite fermionic pairs in the admixture to another excited
molecular bound state $A^{1}\Sigma_{u}^{+}(v'=68)$, which suffers
from the spontaneous-emission loss at the rate $\gamma$. In the case
that this is the dominant loss channel, one can determine the photo-excitation
rate from the loss rate of the system, i.e., $-\dot{N}/N.$ The above
mentioned Fermi gas system can be well described by a two-channel
model Hamiltonian $\mathscr{H}=\mathscr{H}_{0}^{(a)}+\mathcal{\mathscr{H}}_{\textrm{int}}^{(a)}+\mathscr{H}^{(am)}+\mathscr{H}^{(m)}$,
where \citep{Ohashi2002,Falco2005,Liu2005}
\begin{eqnarray}
\mathscr{H}_{0}^{(a)} & = & \sum_{\mathbf{k}\sigma}\left(\epsilon_{\mathbf{k}}-\mu\right)c_{\mathbf{k}\sigma}^{\dagger}c_{\mathbf{k}\sigma},\\
\mathcal{\mathscr{H}}_{\textrm{int}}^{(a)} & = & \frac{u_{0}}{V}\sum_{\mathbf{k}\mathbf{k}'\mathbf{q}}c_{\frac{\mathbf{q}}{2}+\mathbf{k}\uparrow}^{\dagger}c_{\frac{\mathbf{q}}{2}-\mathbf{k}\downarrow}^{\dagger}c_{\frac{\mathbf{q}}{2}-\mathbf{k}'\downarrow}c_{\frac{\mathbf{q}}{2}+\mathbf{k}'\uparrow},\\
\mathscr{H}^{(am)} & = & \frac{g_{0}}{\sqrt{V}}\sum_{\mathbf{kq}}\left[\phi_{g\mathbf{q}}^{\dagger}c_{\frac{\mathbf{q}}{2}-\mathbf{k}\downarrow}c_{\frac{\mathbf{q}}{2}+\mathbf{k}\uparrow}+\textrm{h.c.}\right],\\
\mathscr{H}^{(m)} & = & \sum_{\mathbf{q}}\left(\begin{array}{cc}
\phi_{g\mathbf{q}}^{\dagger} & \phi_{e\mathbf{q}}^{\dagger}\end{array}\right)\mathcal{M}_{\mathbf{q}}\left(\begin{array}{c}
\phi_{g\mathbf{q}}\\
\phi_{e\mathbf{q}}
\end{array}\right),
\end{eqnarray}
describe respectively the kinetic Hamiltonian of atoms, the background
interaction Hamiltonian of atoms with the interaction strength $u_{0}$,
the coupling between atoms and ground-state molecules ($g_{0}$),
and the molecules in both ground and excited molecular states. Here,
$\epsilon_{\mathbf{k}}\equiv\hbar^{2}\mathbf{k}^{2}/(2m)$ is the
free dispersion relation, $\mu$ is the chemical potential of atoms
and, $c_{\mathbf{k}\sigma}$, $\phi_{g\mathbf{q}}$ and $\phi_{e\mathbf{q}}$
are the annihilation field operators of atoms (with spin $\sigma=\uparrow,\downarrow$),
ground-state molecules and excited-state molecules, respectively.
In the last Hamiltonian for molecules, $\mathcal{M}_{\mathbf{q}}$
is a 2 by 2 matrix,
\begin{equation}
\mathcal{M}_{\mathbf{q}}=\left[\begin{array}{cc}
\frac{\epsilon_{\mathbf{q}}}{2}-2\mu+\delta_{g0} & \Omega/2\\
\Omega/2 & \frac{\epsilon_{\mathbf{q}}}{2}-2\mu+\delta_{e0}-\Delta-i\frac{\gamma}{2}
\end{array}\right],
\end{equation}
where $\Omega$ and $\Delta$ are the Rabi coupling and detuning of
the photo-excitation laser, respectively, $\gamma$ is loss rate,
and $\delta_{g0}$ and $\delta_{e0}$ are the bare detunings of the
molecular states.

The subscript ``0'' in various parameters indicates that these \emph{bare}
parameters are to be renormalized and related to some physical observables
\citep{Falco2005,Liu2005}. For example, $u_{0}$ will be expressed
in terms of the background $s$-wave scattering length $a_{bg}$,
$g_{0}$ will be replaced by the width of the Feshbach resonance $W$,
and finally $\delta_{g0}$ and $\delta_{e0}$ will correspond to the
detunings of the closed-channels. The parameter renormalization has
been discussed in detail in the literature \citep{Falco2005,Liu2005}.
It is very easy to implement: we can simply do the following replacements
in \emph{final} equations derived, i.e.,
\begin{eqnarray}
u_{0} & \rightarrow & u=\frac{4\pi\hbar^{2}}{m}a_{bg},\\
g_{0} & \rightarrow & g=\sqrt{\frac{4\pi\hbar^{2}}{m}a_{bg}W\mu_{ag}},\\
\delta_{g0} & \rightarrow & \delta_{g}=\mu_{ag}\left(B-B_{0}\right),\\
\delta_{e0} & \rightarrow & \delta_{e}=\mu_{ae}\left(B-B_{0}\right).
\end{eqnarray}
For $^{6}$Li atoms near the Feshbach resonance at $B_{0}\simeq832.18$
G, $a_{bg}\simeq-1582a_{0}$ ($a_{0}=0.0529177$ nm), $W=-262.3$
G, $\mu_{ag}=2\mu_{a}-\mu_{g}\simeq2\mu_{B}$ ($\mu_{B}$ the Bohr
magneton) is the difference in the magnetic moments of atoms and of
molecules for the ground state \citep{Zurn2013}, and similarly $\mu_{ae}=2\mu_{a}-\mu_{e}\simeq2\mu_{B}$.
For the excited molecular state, the loss rate $\gamma\simeq2\pi\hbar\times11.8$
MHz. The Rabi coupling $\Omega$ is typically at the order of $2\pi\times1.0$
MHz. In addition, to convert the Zeeman field to frequency, we could
use $\mu_{B}\times1\textrm{G}\simeq2\pi\hbar\times1.3996\textrm{MHz}$.

\section{Universal Tan relation for photo-excitation}

According to Tan, Braaten and Platter \citep{Braaten2008}, the loss
rate of a quantum many-body system with short-range inter-particle
interactions at the number of particles $N$ is given by,
\begin{equation}
-\dot{N}\equiv-\frac{dN}{dt}=\frac{\hbar\left[-\textrm{Im}a\left(B\right)\right]}{2\pi m\left|a\left(B\right)\right|^{2}}\mathcal{I}.\label{eq:LossRate}
\end{equation}
Here, $a(B)$ is the $s$-wave scattering length at a given Zeeman
field $B$ and $\mathcal{I}$ is the contact coefficient of the system.
The loss rate equation can be intuitively understood from Tan's adiabatic
relation, 
\begin{equation}
\frac{dE}{d\left(1/a\right)}=-\frac{\hbar^{2}}{4\pi m}\mathcal{I},
\end{equation}
which relates the total energy $E$ of the many-body system to the
contact coefficient $\mathcal{I}$. Due to the coupling to the dissipative
excited molecular state, the total energy acquires a relatively small
imaginary part $E-iE_{\textrm{I}}/2$, where $E_{\textrm{I}}/\hbar=-dN/dt$
can be understood as the loss rate of the system, since the time evolution
of the system can be given by $\exp[-i\left(E-iE_{\textrm{I}}/2\right)t/\hbar]$.
Correspondingly, the scattering length also has a small imaginary
part. From the adiabatic relation, we have,
\begin{equation}
-\frac{E_{\textrm{I}}}{2}\simeq\textrm{Im}\left[dE\right]=-\frac{\hbar^{2}\mathcal{I}}{4\pi m}\textrm{Im}\left[d\left(1/a\right)\right],
\end{equation}
which immediately leads to the loss rate equation (\ref{eq:LossRate}).

Let us now calculate the scattering length $a(B)$ for the experimental
photo-excitation process. Taking into account the molecule-mediated
attraction, the total (un-renormalized) interaction between two atoms
can be written into the form,
\begin{equation}
U\left(\mathbf{q},i\nu_{n}\right)=u_{0}+g_{0}\mathscr{D}_{g}^{(0)}\left(\mathbf{q},i\nu_{n}\right),
\end{equation}
where $\mathscr{D}_{g}^{(0)}(\mathbf{q},i\nu_{n})$ is the non-interacting
Green function of molecules in the \emph{ground-state} channel and
is the 11-component of the 2 by 2 matrix,
\begin{equation}
\mathscr{D}^{(0)}\left(\mathbf{q},i\nu_{n}\right)=\left[i\nu_{n}-\mathcal{M}_{\mathbf{q}}\right]^{-1}.
\end{equation}
By explicitly working out the inverse of the matrix, we obtain that,
\begin{eqnarray}
\mathscr{D}_{g}^{(0)} & \left(\mathbf{q},i\nu_{n}\right)= & \left[i\nu_{n}-\epsilon_{\mathbf{q}}/2+2\mu-\delta_{g0}\right.-\nonumber \\
 &  & \left.\frac{\Omega^{2}/4}{i\nu_{n}-\epsilon_{\mathbf{q}}/2+2\mu-\delta_{e0}+\Delta+i\gamma/2}\right]^{-1}.\label{eq:GF}
\end{eqnarray}
The last term in the bracket gives the conventional Stark shift to
the ground-state molecules due to the coupling to the excited molecular
state. In the vacuum, where we set the chemical potential $\mu=0$,
the $s$-wave scattering length $a(B)$ is then given by,
\begin{align}
\frac{4\pi\hbar^{2}a\left(B\right)}{m} & =u_{0}+g_{0}^{2}\mathscr{D}_{g}^{(0)}\left(\mathbf{0},0\right),\\
 & =u-\frac{g^{2}}{\delta_{g}+\Omega^{2}/\left[4\left(-\delta_{e}+\Delta+i\gamma/2\right)\right]}.
\end{align}
In the last step, we have replaced the un-regularized parameters with
the regularized ones. In the absence of the Rabi coupling, i.e., $\Omega=0$,
it is easy to check that we recover the well-known expression for
the scattering length near a Feshbach resonance,
\begin{equation}
a_{s}\left(B\right)=\frac{m}{4\pi\hbar^{2}}\left[u-\frac{g^{2}}{\delta_{g}}\right]=a_{bg}\left(1-\frac{W}{B-B_{0}}\right).
\end{equation}
Here, the subscript ``$s$'' in $a_{s}(B)$ indicates the $s$-wave
scattering length \emph{without} Rabi coupling. In the presence of
Rabi coupling, instead we would obtain,
\begin{equation}
a\left(B\right)=a_{bg}-\frac{a_{bg}\Gamma}{\delta_{g}+\Omega^{2}/\left[4\left(-\delta_{e}+\Delta+i\gamma/2\right)\right]},
\end{equation}
where $\Gamma\equiv\mu_{ag}W$ is the characteristic energy related
to the width of the Feshbach resonance. It is straightforward to obtain,
\begin{equation}
-\frac{\textrm{Im}a\left(B\right)}{\left|a\left(B\right)\right|^{2}}=\left(\frac{\Gamma}{a_{bg}}\right)\frac{\left[\Omega^{2}/\left(2\gamma\right)\right]}{\left(\delta_{g}-\Gamma\right)^{2}+c^{2}\left[\Omega^{2}/\left(2\gamma\right)\right]^{2}},
\end{equation}
where the coefficient
\begin{equation}
c=1-\frac{4\left(\delta_{e}-\Delta\right)\left(\delta_{g}-\Gamma\right)}{\Omega^{2}}.
\end{equation}
By substituting it into the loss rate expression (\ref{eq:LossRate}),
we find that,
\begin{equation}
-\frac{\dot{N}}{N}=\frac{\hbar\mathcal{I}}{2\pi mN}\left(\frac{\Gamma}{a_{bg}}\right)\frac{\left[\Omega^{2}/\left(2\gamma\right)\right]}{\left(\delta_{g}-\Gamma\right)^{2}+c^{2}\left[\Omega^{2}/\left(2\gamma\right)\right]^{2}}.
\end{equation}
By recalling that
\begin{equation}
\delta_{g}-\Gamma=-\frac{\Gamma}{\left[1-a_{bg}/a_{s}\right]},
\end{equation}
and defining \citep{Werner2009}
\begin{eqnarray}
R_{*} & = & \frac{\hbar^{2}}{ma_{bg}\Gamma},\\
\mathcal{I} & = & 4\pi Nk_{F}\mathcal{F}\left(\frac{1}{k_{F}a_{s}}\right),
\end{eqnarray}
we finally arrive at,
\begin{equation}
-\frac{\dot{N}}{N}=\hbar^{-1}k_{F}R_{*}\mathcal{F}\left(\frac{1}{k_{F}a_{s}}\right)L\left(\Delta\right),\label{eq:LossRate2}
\end{equation}
where the line-shape function of the photo-excitation laser $L(\Delta)$
takes the form,
\begin{equation}
L\left(\Delta\right)=\frac{\Omega^{2}/\gamma}{\left[1-a_{bg}/a_{s}\right]^{-2}+\left[\frac{\Omega^{2}-4\left(\delta_{e}-\Delta\right)\left(\delta_{g}-\Gamma\right)}{2\gamma\Gamma}\right]^{2}}.\label{eq:LineShape}
\end{equation}

\begin{figure}[t]
\begin{centering}
\includegraphics[width=0.98\columnwidth]{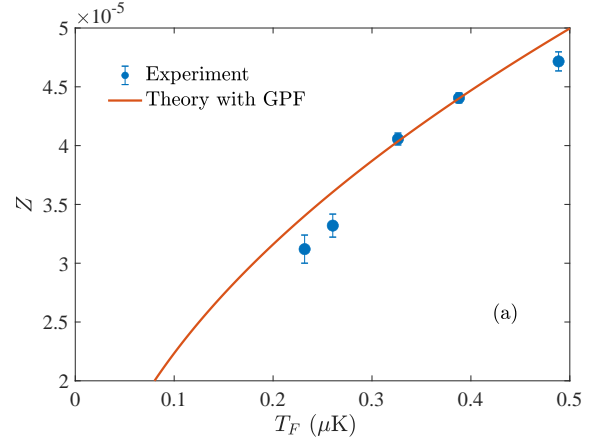}
\par\end{centering}
\begin{centering}
\includegraphics[width=0.98\columnwidth]{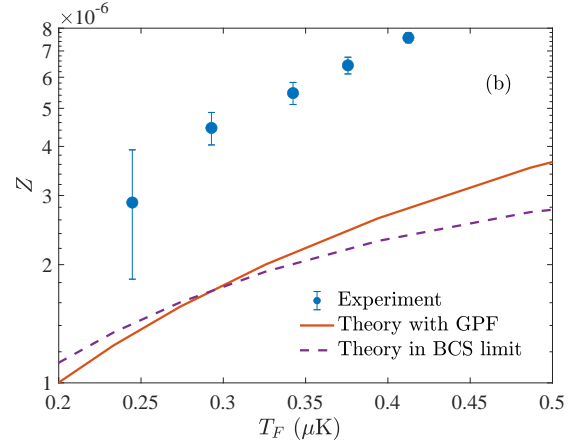}
\par\end{centering}
\caption{\label{fig1_ndep} Density dependence of the closed-channel fraction
$Z$ of a strongly interacting $^{6}$Li Fermi gas in the unitary
limit $B=832$ G (a) and on the BCS side at the magnetic field $B=925$
G. Here, the density is expressed in terms of the Fermi energy $T_{F}$.
The low-temperature data from USTC Shanghai group (symbols) \citep{Liu2019arXiv}
are compared with our zero-temperature theoretical predictions using
the contact obtained from, a Gaussian pair fluctuation theory (solid
line) \citep{Hu2011} and a perturbation theory (dashed line) \citep{Werner2009}.}
\end{figure}

Under the resonant condition $\delta_{e}=\Delta$ and with a weak
laser intensity $\Omega^{2}\ll\gamma\Gamma$, the line-shape function
is $L(\Delta)=(1-a_{bg}/a_{s})^{2}\Omega^{2}/\gamma$. As $\hbar^{-1}\Omega^{2}/\gamma$
is the effective decay rate of the ground molecular state for a resonant
transition, see, i.e., Eq. (\ref{eq:GF}), we may rewrite \citep{Partridge2005,Liu2019arXiv}
\begin{equation}
-\frac{\dot{N}}{N}=Z\frac{\Omega^{2}}{\hbar\gamma},
\end{equation}
where $Z$ is the molecular fraction in the ground molecular state
and hence is given by 
\begin{equation}
Z=k_{F}R_{*}\mathcal{F}\left(\frac{1}{k_{F}a_{s}}\right)\left[1-\frac{a_{bg}}{a_{s}}\right]^{2}=\frac{\mathcal{I}R_{*}}{4\pi N}\left[1-\frac{a_{bg}}{a_{s}}\right]^{2}.\label{eq:Z}
\end{equation}
Thus, in the limit of a weak laser intensity, we recover the universal
relation that links the closed-channel molecular fraction to the contact
coefficient, firstly pointed out by Werner \textsl{et al.} \citep{Werner2009}.
Our loss rate equation Eq. (\ref{eq:LossRate2}) does not have the
weak probe restriction. It holds as long as the assumption of a single
$s$-wave scattering length $a_{s}(B)$ is applicable.

\begin{figure}[t]
\begin{centering}
\includegraphics[width=0.98\columnwidth]{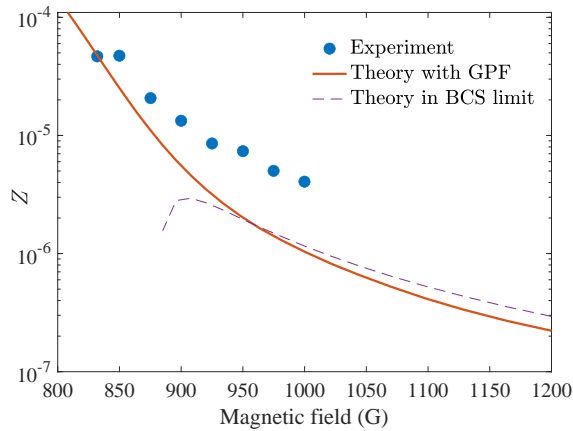}
\par\end{centering}
\caption{\label{fig2_Bdep} Magnetic field dependence of the closed-channel
fraction $Z$ of a strongly interacting $^{6}$Li Fermi gas at the
density characterized by a Fermi temperature $T_{F}=0.45$ $\mu$K.
The low-temperature data from USTC Shanghai group (symbols) \citep{Liu2019arXiv}
are compared with our zero-temperature theoretical predictions using
the contact obtained from, a Gaussian pair fluctuation theory (solid
line) \citep{Hu2011} and a perturbation theory (dashed line) \citep{Werner2009}.}
\end{figure}

\section{Experimental relevance}

\subsection{Photo-excitation measurement at USTC Shanghai}

In a recent experiment at USTC Shanghai \citep{Liu2019arXiv}, the
laser intensity is weak and the Rabi coupling is measured to be $\Omega^{2}/\gamma\simeq2\pi\hbar\times0.136(1)$
MHz. This is much smaller than the characteristic energy of the Feshbach
resonance width, i.e., $\Gamma=\mu_{ag}W=2\mu_{B}\times262.3\textrm{G}\simeq2\pi\hbar\times734\textrm{MHz}$.
Therefore, for resonant excitation ($\delta_{e}=\Delta$) we may use
the weak-intensity result,
\begin{equation}
-\frac{\dot{N}}{N}=k_{F}R_{*}\mathcal{F}\left(\frac{1}{k_{F}a_{s}}\right)\left[1-\frac{a_{bg}}{a_{s}}\right]^{2}\left(\frac{\Omega^{2}}{\hbar\gamma}\right),
\end{equation}
to analyze the experimental data and to calculate the closed-channel
fraction $Z$ in the ground molecular state according to Eq. (\ref{eq:Z}).
In Fig. \ref{fig1_ndep}, we present the density dependence of the
closed-channel fraction $Z$ in the unitary limit $B=832$ G (a) and
on the BCS side $B=925$ G (b), experimentally measured at USTC Shanghai
\citep{Liu2019arXiv} and theoretically calculated by using Eq. (\ref{eq:Z})
with the contact obtained either from a Gaussian pair-fluctuation
theory \citep{Hu2011,Hu2006,Hu2007} or from a perturbation theory
\citep{Werner2009}. The latter is only applicable in the deep BCS
limit. We find a good agreement between theory and experiment in the
unitary limit. However, on the BCS side, there is about a factor of
three difference. In Fig. \ref{fig2_Bdep}, we show the comparison
as a function of the magnetic field. The agreement seems to become
increasingly worse when we increase the magnetic field.

We do not fully understand why there is about a factor of three difference
on the BCS side. A possible source for the discrepancy is the Rabi
coupling $\Omega$, which is only experimentally calibrated on the
BEC side and is then assumed to be \emph{invariant} across the BEC-BCS
crossover. Mathematically, the Rabi coupling is given by the overlap
between the two wave-functions of the bound molecular states \citep{Partridge2005},
i.e., 
\begin{equation}
\Omega=\left\langle \psi_{v'=68}(S=0)\left|\vec{d}\cdot\vec{E}_{\textrm{L}}\right|\psi_{v=38}(S=0)\right\rangle ,
\end{equation}
where $\vec{d}$ is the transition dipole and $\vec{E}_{\textrm{L}}$
is the laser field for photon excitation. This wave-function overlap
could change notably on the BCS side, as the ground molecular state
becomes increasingly affected by the admixture with atoms. A few-body
calculation is needed, i.e., following the theoretical photoassociation
work \citep{Pellegrini2008}, in order to fully understand the magnetic
dependence of the Rabi coupling near a Feshbach resonance. 

\subsection{Contact measurement at finite temperature}

To avoid the complications due to the magnetic field dependence in
the Rabi coupling, we may perform the photo-excitation measurement
at a fixed magnetic field and consider different temperatures. To
calibrate $\Omega^{2}/\gamma$, we can go to the strong laser intensity
regime, in the sense that $\Omega^{2}/\gamma$ can be enlarged to
be comparable to $\Gamma$, as long as there is no significant heating.
This will considerably increase the experimental resolution for the
loss rate measurement. 

In greater detail, by increasing the laser intensity $I$ to increase
the Rabi coupling (i.e., $\Omega^{2}=\alpha I$), under the resonant
condition we may first confirm the predicted line shape in Eq. (\ref{eq:LineShape}),
\begin{equation}
L\left(\Delta\right)=2\Gamma\frac{\Omega^{2}/\left(2\gamma\Gamma\right)}{\left[1-a_{bg}/a_{s}\right]^{-2}+\left[\Omega^{2}/\left(2\gamma\Gamma\right)\right]^{2}},\label{eq:LineShapeResonant}
\end{equation}
which takes a maximum $2\Gamma(1-a_{bg}/a_{s})$. At a given magnetic
field, the detunings $\Delta$, $\delta_{g}$, $\delta_{e}$ and $a_{s}(B)$
are known precisely and the zero-temperature contact has been also
determined to a reasonable accuracy. Using these knowledge, we can
determine the proportional factor $\alpha$ and hence calibrate the
Rabi coupling $\Omega$. Then, working at a fixed Rabi coupling, we
may tune the temperature of the system and consequently measure the
temperature dependence of the contact coefficient. As the contact
coefficient changes significantly across the superfluid phase transition,
we can ultimately determine the critical temperature of a strongly
interacting Fermi gas across the BEC-BCS crossover.

\section{Conclusions}

In summary, we have derived a new universal relation for the photon-excitation
measurement of a strongly interacting Fermi gas near a Feshbach resonance.
We have shown that the determination of the photo-excitation rate
can directly give Tan's contact coefficient. In the limit of a weak
laser intensity, our relation reduces to the well-known relation for
the closed-channel molecular fraction, derived earlier by Werner,
Tarruell and Castin \citep{Werner2009}. With a strong laser intensity,
we anticipate that the photon-excitation measurement can have improved
experimental resolution and hence provide an accurate way to measure
the temperature dependence of the contact coefficient and also the
critical temperature at the BEC-BCS crossover.
\begin{acknowledgments}
We thank Yu-Ao Chen, Qijin Chen, Xing-Can Yao, and Johannes Hecker
Denschlag for the simulating discussions. This research was supported
by the Australian Research Council's (ARC) Discovery Program, Grants
No. DE180100592 and No. DP190100815 (J.W.), Grant No. DP170104008
(H.H.), and Grant No. DP180102018 (X.-J.L).
\end{acknowledgments}

\end{document}